\documentstyle[prl,aps,preprint]{revtex}

\begin{document}

\newcommand{\be}{\begin{equation}}
\newcommand{\ee}{\end{equation}}
\newcommand{\bq}{\begin{eqnarray}}
\newcommand{\eq}{\end{eqnarray}}

\setcounter{page}{0}
\def\footnoterule{\kern-3pt \hrule width\hsize \kern3pt}
\tighten
\title{
Virasoro Algebra and $O(N)$ $\sigma $-model
}
\author{Jiannis Pachos
\footnote{Email address: {\tt pachos@ctp.mit.edu}}}

\address{Center for Theoretical Physics \\
Massachusetts Institute of Technology \\
Cambridge, Massachusetts 02139 \\
{~}}

\date{MIT-CTP-2633, May 1997}
\maketitle

\thispagestyle{empty}

\begin{abstract}

Within the framework of a local expansion of the logarithm of the
$O(N)$ $\sigma$-model vacuum functional, valid for slowly varying fields, 
the modified Virasoro algebra is studied. The operator-like central charge 
term is given, up to second order, for a general functional.

\end{abstract}

\vspace*{\fill}

\pacs{xxxxxx}

\section{Introduction}

During the last twenty years a great deal of work has been done on
conformally symmetric theories. Importantly, they can be exactly solved to
give critical exponents of two dimensional theories, aiding to their
classification. Moreover, their conformal symmetry enables the association
of strongly interacting fields with weakly coupled ones, which are easy
to elaborate. To make physically interesting theories out of them it is
necessary to incorporate interactions (e.g. curvature) in the free case. The
cases of interacting fields which preserve the conformal symmetry in a
stronger or a weaker sense have been studied in the literature \cite{Green},
as well as theories where their interactions destroy this symmetry. One of
the latter is the non-linear $O(N)$ $\sigma $-model, where
the $O(N)$ symmetry generates a mass term
in the quantum level, destroying the classical conformal symmetry of the
model.

A considerable amount of interest is concentrated on the cylindrical
space-time ${\bf R}^1\times S^1$, which shares many features with string
theory \cite{jackiw2}. We will face one of them, the Virasoro algebra, which
has been previously studied through different quantization procedures, by
using the functional formalism.
Our aim is to set up a general formalism for the study of a modified form of
the Virasoro algebra for the $O(N)$ $\sigma $-model, for which 
instead of the usual
central charge term we expect operator-like terms as a quantum anomaly
extension of this algebra. As the form of the vacuum functional is 
not known, we will calculate the anomaly extension up to an undefined 
function on which the operator-like term acts. We expect that further study 
of the renormalization group properties of the model \cite{anselmi}, 
will determine completely this term up to the approximation scheme
adopted in the following. Note that in this scheme we expand the vacuum
functional for configurations with small momentum
(i.e. large distances), where phenomena of mass generation are dominant.

\section{Virasoro Algebra}

Let us briefly review the Virasoro Algebra for string theory.
We can have a string on an $N$ dimensional manifold 
parameterized by $X^\mu $ while the string
spans a $1+1$ dimensional Minkowski world sheet with coordinates $\sigma $
and $\tau $. The position of the string on the manifold is given by the
functions $X^\mu (\sigma ,\tau )$. The string can be open or closed. In the
second case we demand $X(\sigma ,\tau )$ to be periodic in $\sigma $, which
we assume runs around the string in the interval $[-\pi ,\pi ]$, while
in the first case $\sigma$ takes values in the interval $[0,\pi]$. The action
of the string can be written 
\begin{equation}
\label{string}S=-\frac 1{2\pi }\int d\sigma d\tau \,\eta ^{\alpha \beta
}\partial _\alpha X\cdot \partial _\beta X\,\, ,
\end{equation}
which is the action for a $\sigma $-model, now chosen to be in a flat
background manifold. 
In the Hamiltonian formalism $P=\dot X$ ($\cdot $ stands for a derivative 
with respect to time) and at the quantum level the action 
(\ref{string}) is accompanied by the commutation relations 
\begin{equation}
\label{coca}[X^\mu (\sigma ),P_\nu (\sigma ^{\prime })]=i\delta _\nu ^\mu
\delta (\sigma -\sigma ^{\prime })\,\, , \,\,\,\,\,\,\,
[X^\mu (\sigma ),X_\nu (\sigma ^{\prime })]=
[P^\mu (\sigma ),P_\nu (\sigma ^{\prime })]=0
\end{equation}
For the case of the closed string there are two oscillating modes going
``left'' and ``right''. These modes can be generated by the Virasoro
operators%
$$
L_n=\frac 14\int\limits_{-\pi }^\pi d\sigma :(P^\mu -X^{\prime \mu })(P_\mu
-X_\mu ^{\prime }):e^{-in\sigma } 
$$
$$
\widetilde{L}_m=\frac 14\int\limits_{-\pi }^\pi d\sigma :(P^\mu +X^{\prime
\mu })(P_\mu +X_\mu ^{\prime }):e^{im\sigma } 
$$
for $m$ and $n$ integers different in general, that is the two modes are
independent from each other. The operators%
\be
\alpha _n^\mu \equiv \frac 1{\sqrt{\pi }}\int d\sigma\, e^{-in\sigma }\left(
P^\mu (\sigma )-X^{\prime \mu }(\sigma )\right) 
\,\,\,\,\, \text {and} \,\,\,\,\,
\widetilde{\alpha }_n^\mu \equiv \frac 1{\sqrt{\pi }}\int d\sigma \,
e^{in\sigma }\left( P^\mu (\sigma )+X^{\prime \mu }(\sigma )\right) 
\ee
play the role of the creation operators for $n<0$ and the annihilation ones
for $n>0$ with respect to the vacuum state, $|0\rangle $, of the string. 
$\alpha $ and $ \widetilde{\alpha }$ satisfy the commutation relations
\be
[\alpha _n^\mu ,\alpha _m^\nu ]=\eta ^{\mu \nu }n\delta _{m+n,0}=[\widetilde{%
\alpha }_n^\mu ,\widetilde{\alpha }_m^\nu ] 
\,\, ,\,\,\,\,\,\,\,\,\,\,
[\alpha _n^\mu ,\widetilde{\alpha }_m^\nu ]=0 
\ee
following from (\ref{coca}). We can re-write the Virasoro operators as 
\begin{equation}
\label{vir1}L_n=\frac 12\sum :\alpha _{n+p}^\mu \alpha _{-p}^\nu :\eta _{\mu
\nu } \,\,\,\,\,\,
\text {and}\,\,\,\,\,\,
\widetilde{L}_n=\frac 12\sum :\widetilde{\alpha }_{n+p}^\mu 
\widetilde{\alpha }_{-p}^\nu :\eta _{\mu \nu } \,\, ,
\end{equation}
where now the meaning of the normal ordering symbol $::$ is clearly defined
with respect to the operators $\alpha $ and $\widetilde{\alpha }$ acting 
on the vacuum state, $|0\rangle$. It is
actually needed only for the case $n=0$, as $L_0+\widetilde{L}_0+2$
represents the Hamiltonian of the closed string and needs normalization of
the vacuum energy (performed here by normal ordering). We can check that the
normal ordered Virasoro operators satisfy the well-known Virasoro algebra 
\begin{equation}
\label{viro}\left[ L_n,L_m\right] =(n-m)L_{n+m}+\frac N{12}\delta
_{n,-m}(n^3-n)\,\, . 
\end{equation}
The same algebra is also satisfied by the $\widetilde{L}$ operators,
and by the Virasoro operators of the open string. In \cite{bollini},
an algebraic derivation of (\ref{viro}) is presented.

\section{The Vacuum State of the String}

We can construct the wave functional $\langle X|0\rangle $, that represents
the vacuum state $|0\rangle $, explicitly by using the annihilation relation
$\alpha _{n \mu} |0\rangle =0$
for $n>0$. The equation it has to satisfy is
\begin{equation}
\label{gaus}
\alpha _{n\mu }\langle X|0\rangle =0\,\,\,\Rightarrow\,\,\, 
\frac 1{\sqrt{\pi }}\int d\sigma \, e^{-in\sigma }\left( -i\frac \delta {\delta
X^\mu (\sigma )}-X_\mu ^{\prime }(\sigma )\right) \langle X|0\rangle
=0\,\,,
\end{equation}
where the momentum operator has been represented by 
$P_{\mu}(\sigma)=-i{\delta \over {\delta X^\mu(\sigma )}}$.
From (\ref{gaus}) we see that the vacuum state will have a Gaussian
localized form. Let us take%
$$
\langle X|0\rangle =
\exp \left(\int \!\! \int d\sigma d\sigma ^{\prime } \,X^\mu (\sigma
)H(\sigma ,\sigma ^{\prime })X_\mu (\sigma ^{\prime })\right) 
$$
where $H(\sigma ,\sigma ^{\prime })$ is a symmetric function of $\sigma $
and $\sigma ^{\prime }$, to be calculated. As all the points on the closed
string are equivalent, $H$ should be a function of the difference $\sigma
-\sigma ^{\prime }$. So its decomposition in modes will be%
$$
H(\sigma ,\sigma ^{\prime })=H(\sigma -\sigma ^{\prime
})=\sum\limits_{m=-\infty }^\infty H^me^{im(\sigma -\sigma ^{\prime })} 
$$
where $H^m$ are the Fourier components of $H(\sigma ,\sigma ^{\prime })$.
After substituting into (\ref{gaus}) we obtain
$H^m=-\frac m{4\pi }$
for $m\geq 0$. As $H(\sigma ,\sigma ^{\prime })$ has to be symmetric in its
arguments we finally have%
$$
H^m=-\frac{|m|}{4\pi },\,\,\,\,\,\,\,\,\,\,\,\,
\text{for any integer }m \,\, . 
$$

\section{Functional Method for the Central Charge Term}

We can define the regularized Virasoro operators as
\be
\label{gv}
L[u]=\frac 14\int\!\!\int d\sigma d\sigma ^{\prime } \,u(\sigma ,\sigma ^{\prime
})G^{\mu \nu }_s(\sigma ,\sigma ^{\prime })
\big[P_\mu (\sigma )-X_\mu ^{^{\prime}}(\sigma )\big] 
\big[P_\nu (\sigma ^{\prime })-X_\nu ^{\prime }(\sigma ^{\prime
}) \big]\equiv 
\ee
$$
\frac 14\int\!\!\int d\sigma d\sigma ^{\prime }\,u(\sigma ,\sigma ^{\prime })
G^{\mu \nu }_s(\sigma ,\sigma ^{\prime })
R_\mu(\sigma )R_\nu (\sigma ^{\prime }) \,\, ,
$$
where $G_s^{\mu \nu }(\sigma ,\sigma ^{\prime })$ is a Kernel to point split
the double action of the functional differentiations, satisfying the
condition $\lim _{s\rightarrow 0}G_s^{\mu \nu }(\sigma ,\sigma ^{\prime
})=\eta ^{\mu \nu }\delta (\sigma ,\sigma ^{\prime })$, $R_\mu (\sigma
)\equiv $ $P_\mu (\sigma )-X_\mu ^{^{\prime }}(\sigma )$, $u(\sigma
,\sigma ^{\prime })$ is the component of a vector field on the circle, $S^1$%
, on which $X(\sigma )$ is defined and is symmetric in $\sigma $ and $\sigma
^{\prime }$. For $u(\sigma ,\sigma
^{\prime })=1$ we get the divergent quantity $L[1]$, which is equivalent to
the divergent Hamiltonian of the string.

We will calculate the commutator $[L[u],L[v]]$ acting on the vacuum state 
$\langle X|0\rangle $ given in the previous section. We have
$$
\left[ L[u],L[v]\right] \langle X|0\rangle = 
$$
$$
\frac 1{16}\int\!\!\int d\sigma d\sigma ^{\prime } \,
d\bar \sigma d\bar \sigma ^{\prime
}u(\sigma ,\sigma ^{\prime })v(\bar \sigma ,\bar \sigma ^{\prime })G^{\mu
\nu }_s(\sigma ,\sigma ^{\prime })G^{\kappa \lambda }_s(\bar \sigma ,\bar \sigma
^{\prime })\times  
$$
$$
\left[ R_\mu (\sigma )R_\nu (\sigma ^{\prime }),R_\kappa (\bar \sigma
)R_\lambda (\bar \sigma ^{\prime })\right] \langle X|0\rangle = 
$$
$$
-\frac 14\int\!\!\int d\sigma d\sigma ^{\prime }\,
d\bar \sigma d\bar \sigma ^{\prime
}u(\sigma ,\sigma ^{\prime })v(\bar \sigma ,\bar \sigma ^{\prime })G^{\mu
\nu }_s(\sigma ,\sigma ^{\prime })G^{\kappa \lambda }_s(\bar \sigma ,\bar \sigma
^{\prime })\times  
$$
\begin{equation}
\label{more}i\eta _{\mu \kappa }\delta ^{\prime }(\sigma ,\bar \sigma
)\left\{ R_\nu (\sigma ^{\prime })R_\lambda (\bar \sigma ^{\prime
})+R_\lambda (\bar \sigma ^{\prime })R_\nu (\sigma ^{\prime })\right\}
\langle X|0\rangle \,\, .
\end{equation}
(\ref{more}) results after applying the relation $\left[ R_\mu (\sigma
),R_\kappa (\bar \sigma )\right] =-2i\delta ^{\prime }(\sigma ,\bar \sigma
)\eta _{\mu \kappa }$ and re-arranging the $\sigma $ variables as well as
the indices. By commuting $R_\lambda (\bar \sigma ^{\prime })$ and $R_\nu
(\sigma ^{\prime })$ and integrating by parts, we get
$$
\left[ L[u],L[v]\right] \langle X|0\rangle = 
$$
$$
\frac i2\int\!\!\int d\sigma d\sigma ^{\prime }\,\left\{ 
\frac \partial {\partial \sigma }u(\sigma ,\sigma ^{\prime
})v(\sigma ,\sigma )G^{\lambda \nu }_s(\sigma,\sigma ^{\prime })
+\right.
\,\,\,\,\,\,\,\,\,\,\,\,\,\,\,\,\,\,\,\,\,\,\,\,\,\,\,\,\,\,\, 
$$
\begin{equation}
\label{kati}\,\,\,\,\,\,\,\,\,\,\,\,\,\,\,\,\,\,\,\,\,\,\,\,\,\,\,\,\,\,\,\,%
\left. 
u(\sigma ,\sigma ^{\prime
})v(\sigma ,\sigma )\frac \partial {\partial \sigma }G^{\lambda \nu }_s(\sigma
,\sigma ^{\prime })
\right\} R_\nu (\sigma
^{\prime })R_\lambda (\sigma )\langle X|0\rangle 
\end{equation}
where the rest of the terms give zero, as they appear symmetric in $u$ and $v$
or include the quantity $\left. \frac \partial {\partial \sigma }G^{\mu \nu
}_s(\sigma ,\sigma ^{\prime })\right| _{\sigma ^{\prime }=\sigma }$, which is
zero. Applying the combination $R_\nu (\sigma ^{\prime
})R_\lambda (\sigma )$ on the vacuum state $\langle X|0\rangle $ the only
ambiguity would be from the action of the two functional
derivatives. That is
$$
\frac \delta {\delta X^\mu (\sigma ^{\prime })}\frac \delta {\delta X^\nu
(\sigma )}\langle X|0\rangle = 
$$
\begin{equation}
\label{derss}\left[ 4\int H(\sigma ^{\prime },\sigma ^{\prime \prime })X_\mu
(\sigma ^{\prime \prime })d\sigma ^{\prime \prime }\int H(\sigma ,\sigma
^{\prime \prime })X_\nu (\sigma ^{\prime \prime })d\sigma ^{\prime \prime
}+2H(\sigma ,\sigma ^{\prime })\eta _{\mu \nu }\right] \langle X|0\rangle\,\,.
\end{equation}
As the Kernel acts on (\ref{derss}) the only divergency will come from the
last term in the square brackets. The other terms will be combined as in the
normal ordered case to give the normalized Virasoro operator appearing on the
r.h.s. of the algebra. Let us study the term
$$
T\equiv -i\int\!\!\int d\sigma d\sigma ^{\prime }\, \left\{ 
\frac \partial {\partial \sigma }u(\sigma ,\sigma ^{\prime })v(\sigma
,\sigma )G^{\mu \nu }_s(\sigma ,\sigma ^{\prime }) + \right.
\,\,\,\,\,\,\,\,\,\,\,\,\,\,\,\,\,\,\,\,\,\,\,\,\,\,\,\,\,\,\, 
$$
\begin{equation}
\label{mid2}\,\,\,\,\,\,\,\,\,\,\,\,\,\,\,\,\,\,\,\,\,\,\,\,\,\,\,\,\,\,\,
\left. 
u(\sigma ,\sigma ^{\prime
})v(\sigma ,\sigma )\frac \partial {\partial \sigma }G^{\lambda \nu }_s(\sigma
,\sigma ^{\prime })
\right\} \eta _{\lambda \nu
}H(\sigma ,\sigma ^{\prime }) \,\, .
\end{equation}
We can take the Kernel to be of the form 
\begin{equation}
\label{keke}G^{\mu \nu }_s(\sigma ,\sigma ^{\prime })=\eta ^{\mu \nu }{\cal G}
_s(\sigma ,\sigma ^{\prime })=\eta ^{\mu \nu }\frac{e^{-(\sigma -\sigma
^{\prime })^2/4s}}{\sqrt{2\pi s}} \,\, ,
\end{equation}
where 
$$
\lim _{s\rightarrow 0}\frac{e^{-(\sigma -\sigma ^{\prime })^2/4s}}{\sqrt{%
2\pi s}}=\delta (\sigma -\sigma ^{\prime }) \,\, . 
$$
Expression (\ref{mid2}) can be symmetrized with respect to $\sigma $ and $%
\sigma ^{\prime }$ so that the summation in $H$ can be re-written as%
$$
H(\sigma ,\sigma ^{\prime })=-\frac 1{4\pi }\sum_{m=-\infty }^\infty
|m|e^{im(\sigma -\sigma ^{\prime })}\rightarrow  
$$
$$
\rightarrow -\frac 2{4\pi }\sum_{m=1}^\infty me^{im(\sigma -\sigma ^{\prime
})}=-\frac 1{2\pi i}\sum_{m=1}^\infty \frac \partial {\partial \sigma
}e^{im(\sigma -\sigma ^{\prime })}=-\frac 1{2\pi i}\frac \partial {\partial
\sigma }\left( \frac 1{1-e^{i(\sigma -\sigma ^{\prime })}}\right)  
$$
Substituting this into (\ref{mid2}) we get
$$
T=-\frac{iN}{2}\int \!\!\int d\sigma d\sigma ^{\prime }\,\left\{ 
\frac{\partial }{\partial \sigma }u(\sigma ,\sigma
^{\prime })v(\sigma ,\sigma ){\cal G}_s(\sigma ,\sigma ^{\prime }) + \right. 
$$
\begin{equation}
\label{mid3}\left. 
u(\sigma ,\sigma ^{\prime
})v(\sigma ,\sigma )\frac{\partial }{\partial \sigma }{\cal G}_s(\sigma
,\sigma ^{\prime })\right\}  
\frac{-1}{2\pi i}\frac{\partial }{\partial \sigma }
\left\{\frac 1{1-e^{i(\sigma -\sigma ^{\prime })}}-
\frac 1{1-e^{-i(\sigma -\sigma ^{\prime })}} \right\}\,\, .
\end{equation}
By integrating by parts the $\sigma$ derivative acting on the kernel, we get
\bq
T=&&
\frac {N}{4 \pi}\int\!\!\int d\sigma d\sigma ^{\prime }\,
2 \frac{\partial }{\partial \sigma }u(\sigma ,\sigma
^{\prime })v(\sigma ,\sigma ){\cal G}_s(\sigma ,\sigma ^{\prime })
\frac{\partial }{\partial \sigma }
\left\{\frac 1{1-e^{i(\sigma -\sigma ^{\prime })}}-
\frac 1{1-e^{-i(\sigma -\sigma ^{\prime })}} \right\}-
\nonumber\\
&&
- \frac{N}{4 \pi} \int\!\!\int d\sigma d\sigma ^{\prime }\,
u(\sigma ,\sigma ^{\prime})v(\sigma ,\sigma )
{\cal G}_s(\sigma,\sigma ^{\prime })
\frac{\partial^2 }{\partial \sigma^2 }
\left\{\frac 1{1-e^{i(\sigma -\sigma ^{\prime })}}-
\frac 1{1-e^{-i(\sigma -\sigma ^{\prime })}} \right\} \,\, .
\label{gourou}
\eq
We can define $x=\sigma-\sigma^\prime$.
In (\ref{gourou}) as $s$ gets small the kernel, ${\cal G} _s (x)$, 
becomes nonzero only for small $x$,
so that we can expand the ratio as
$$
\frac 1{1-e^{i x}}=\frac i{x}+\frac 12-\frac i{12}x+...\,. 
$$
Assuming the $\sigma$ dependence of $u$ to be of the form 
$u(\sigma,\sigma^\prime)$$\equiv$$u({{\sigma+\sigma^\prime} \over 2})$, 
we can make the following expansions
\bq
u(\sigma,\sigma^\prime)=u(\sigma-\frac x2)= 
u(\sigma)-\frac{\partial u(\sigma) }{\partial \sigma} \frac x2+
\frac{\partial^2 u(\sigma)}{\partial \sigma^2 } \frac {x^2}{2^2\,2!}+
\frac{\partial^3 u(\sigma)}{\partial \sigma^3 } \frac {x^3}{2^3\,3!}+...
\eq
and
\bq
\label{par}
\frac{\partial }{\partial \sigma }u(\sigma,\sigma^\prime)=
\frac 12 \frac{\partial u(\sigma)}{\partial \sigma} -
\frac{\partial^2 u(\sigma)}{\partial \sigma^2 } \frac {x}{2^2}+
\frac{\partial^3 u(\sigma)}{\partial \sigma^3 } \frac{x^2}{2^4}+...\,\, .
\eq
At the limit $x \rightarrow 0$ the significant nonzero terms in (\ref{gourou}) are
\bq
&&
T=\frac{N}{4 \pi} \int\!\!\int d\sigma dx \, v(\sigma,\sigma) {\cal G} _s (x)  \left\{ 
\frac{\partial u(\sigma)}{\partial \sigma }\left(- \frac {2i}{x^2}\right) +
\frac{\partial u(\sigma)}{\partial \sigma }\left(- \frac {i}{6}\right)+\right.
\nonumber\\
&&
\left. \frac 1{2^3}\frac{\partial^3 u(\sigma)}{\partial \sigma^3 }(-2i)+
\left(-\frac 12 \right)\frac{\partial u(\sigma)}{\partial \sigma}
\left(-\frac {4i}{x^2} \right)-
\frac{\partial^3 u(\sigma)}{\partial \sigma^3 } \frac {-4i}{2^3\,3!}+... \right\}=
\nonumber\\
&&
-\frac{Ni}{24 \pi}\int d\sigma \left\{\frac{\partial u(\sigma)}{\partial \sigma }
+ \frac{\partial^3 u(\sigma)}{\partial \sigma^3 } \right\} v(\sigma) \,\, .
\eq
We can see that the rest of the terms in (\ref{kati}), apart from $T$,
construct the renormalized part of
$$
\left.
i\int\!\!\int d\sigma d\sigma ^{\prime } \,\frac \partial {\partial \sigma
}u(\sigma ,\sigma ^{\prime })v(\sigma ,\sigma )G^{\nu \lambda }_s
(\sigma ,\sigma^{\prime })
R_\nu (\sigma ^{\prime })R_\lambda (\sigma )\langle X|0\rangle 
\right|_{renorm}
$$
and as we can antisymmetrize in $u$ and $v$ we have
$$
\frac i2\int\!\!\int d\sigma d\sigma ^{\prime }\, 
\left\{ \frac \partial {\partial
\sigma }u(\sigma ,\sigma ^{\prime })v(\sigma ,\sigma )-u(\sigma ,\sigma
)\frac \partial {\partial \sigma }v(\sigma ,\sigma ^{\prime })\right\}
\times 
$$
$$
\left.
G^{\nu \lambda }_s(\sigma ,\sigma ^{\prime })R_\nu (\sigma ^{\prime })R_\lambda
(\sigma )\langle X|0\rangle 
\right|_{renorm}= 
$$

$$
\left.
-iL[[u,v]]\langle X|0\rangle 
\right|_{renorm}
$$
for $[u,u]=uv^{\prime }-u^{\prime }v$, by using (\ref{par}). 
This is the desired result as it is
calculated in \cite{mana1} with another method. We see that the commutator 
$[L[u],L[v]]$ gives the renormalized part of $-iL[[u,v]]$ 
plus a finite constant without any infinities appearing in.
For $u=e^{-in\sigma }$ and $
v=e^{-im\sigma }$ it gives (\ref{viro}).

\section{Applications to the $O(N)$ $\sigma $-model}

A similar treatment can be applied to the $O(N)$ $\sigma $-model, where 
we expect the corresponding algebra to differ from the Virasoro algebra, 
because the $\beta_g$ function is nonzero. In other words this model 
quantized is not conformally invariant. Operators defined in a 
similar way as the Virasoro ones for a conformal theory will satisfy an 
algebra with a different central charge term in structure (e.g. it 
could be a function rather than a constant) and in origin, as 
these operators are no longer generators of a conformal 
transformation. Now $\sigma$ parameterizes the infinite space-line, 
so there will be ``left'' and ``right'' modes, 
as in the case of the closed string. We will
use the form (\ref{gv}) for our generalized Virasoro operators, where
here the coordinates $X(\sigma )\equiv z(\sigma )$ represent a manifold
with $O(N)$ symmetry (see \cite{PJ} for conventions and definitions of the
$O(N)$ $\sigma$-model). 
We can work with operators which have similar form to
the Hamiltonian, Lorentz and momentum operator. By studying their algebra we
can connect them with the generalized Virasoro operators and deduce the
algebra the latter satisfy. Instead of a general vector $u(\sigma )$, we can
use powers of the $\sigma $ variable (i.e. $\sigma ^k$, $k\geq 0$). 
For a covariant functional derivative given from
\begin{equation}
\frac D{Dz^\mu (\sigma )} V^{\nu}(\sigma^\prime)
={\frac{\delta V^{\nu}(\sigma^{\prime})}
{\delta z^{\mu}(\sigma)}}+\int d\sigma^{\prime \prime}\,
{\bf \Gamma }_{\mu \rho}^{\nu}(\sigma,\sigma^{\prime},\sigma^{\prime\prime})
V^{\rho}(\sigma^{\prime\prime}) \,\, , 
\end{equation}
where ${\bf \Gamma }_{\mu\rho}^{\nu}
(\sigma,\sigma^{\prime},\sigma^{\prime \prime})=\delta
(\sigma-\sigma^{\prime})\,\delta (\sigma^{\prime}-\sigma^{\prime \prime})
\,{\Gamma }_{\mu\rho}^{\nu}(z(\sigma))$,
when e.g. acting on a local vector $V^{\mu}(\sigma^{\prime})$, 
we can make the following definitions
$$
P\equiv -i\int\!\!\int d\sigma d\sigma ^{\prime } \, 
G^{\mu \nu }_s(\sigma ,\sigma
^{\prime })g_{\nu \kappa }z^{\prime \kappa }\frac D{Dz^\mu (\sigma )}\equiv
-i\int d\sigma {\cal P}(\sigma ) 
$$
$$
H\equiv \frac 12\left[ -\int\!\!\int d\sigma d\sigma ^{\prime }\,
G^{\mu \nu }_s(\sigma
,\sigma ^{\prime })\frac D{Dz^\mu (\sigma )}\frac D{Dz^\nu (\sigma ^{\prime
})}+\int d\sigma g_{\mu \nu }(\sigma )z^{\prime \mu }z^{\prime \nu }\right]
\equiv \frac 12\int d\sigma {\cal H} 
$$
$$
L^k\equiv \frac 12\left[ -\int\!\!\int d\sigma d\sigma ^{\prime } \,
\left( \frac{\sigma
+\sigma ^{\prime }}2\right) ^kG^{\mu \nu }_s(\sigma ,\sigma ^{\prime })\frac
D{Dz^\mu (\sigma )}\frac D{Dz^\nu (\sigma ^{\prime })}+\right. 
$$
$$
\left. \int\!\!\int d\sigma d\sigma ^{\prime } \,
\left( \frac{\sigma +\sigma ^{\prime }}
2\right) ^kg_{\mu \nu }(\sigma )z^{\prime \mu }z^{\prime \nu }\right] \equiv
M^k+\widetilde{{\cal B}} \equiv \frac 12\int d\sigma \, \sigma ^k{\cal H} 
$$
$$
{\cal B}\equiv \int d\sigma \,g_{\mu \nu }(\sigma )z^{\prime \mu }z^{\prime
\nu }\,\,\,\,\text{and}\,\,\,\,\,\widetilde{{\cal B}}\equiv \int d\sigma\,
\sigma ^kg_{\mu \nu }(\sigma )z^{\prime \mu }z^{\prime \nu } \,\, ,
$$
where $\,\widetilde{{\cal B}}$ could be described with two $\sigma $
integrations connected by the Kernel $G$ instead of the metric $g$ and the
term $\sigma ^k$ written as $(\sigma +\sigma ^{\prime })^k/2^k$. However, this
will only differ from the given expression by terms of order $O(s^n)$ with $
n>0$, which will vanish when the limit $s\rightarrow 0$ is taken. 
The kernel $G_s^{\mu\nu}(\sigma, \sigma^{\prime})\equiv{\cal G}
_s(\sigma,\sigma^{\prime})\,{\ K}^{\mu\nu}
(\sigma,\sigma^{\prime};s)$, where 
${\ K}^{\mu\nu}(\sigma,\sigma^{\prime};s)$, with 
${\ K}^{\mu\nu}(\sigma,\sigma;0)=g^{\mu \nu}(\sigma)$, is a function 
of $z$ and its form can be derived after considering rotational symmetry
and Poincar\'e invariance (see \cite{PJ}, \cite{Jian} for its form up to
second order). We can take, as in \cite{Jian}, 
${\cal G}_s(\sigma,\sigma)={\cal G}_s(0)= \frac {b_0^0}{\sqrt{s}}$, 
${\cal G}^{\prime\prime }_s(0)= \frac {b_0^1}{\sqrt{s}^3}$ and
${\cal G}^{\prime \prime \prime \prime}_s (0)= \frac {b_0^2}{\sqrt{s}^5}$.
Now the commutator of $H$ and $L^k$ is
$$
\left[ \int d\sigma\, {\cal H},\int d\sigma\, \sigma ^k{\cal H}\right] \Psi = 
$$
$$
\left[ -\int d\sigma\, \bar \Delta (\sigma )+{\cal B},-\int d\sigma\, \sigma
^k\bar \Delta (\sigma )+\widetilde{{\cal B}}\right] \Psi = 
$$
\begin{equation}
\label{vise}
\left[ -\int d\sigma\, \bar \Delta ,-\int d\sigma\, \sigma ^k\bar
\Delta \right] \Psi +\left[ {\cal B},-\int d\sigma\, \sigma ^k\bar \Delta
\right] \Psi +\left[ -\int d\sigma\, \bar \Delta ,\widetilde{{\cal B}}\right]
\Psi \,\, .
\end{equation}
The second term becomes
$$
\left[ {\cal B},-\int d\sigma\, \sigma ^k\bar \Delta \right] \Psi = 
$$
$$
\int d\sigma\, \sigma ^k\left( (\bar \Delta {\cal \,B})\Psi +2G^{\mu \nu
}_s\frac D{Dz^\mu }{\cal B}\frac D{Dz^\nu }\Psi \right) = 
$$
$$
k(k-1)\frac{b_0^0}{\sqrt{s}}N\int d\sigma\, \sigma ^{k-2}-2N\frac{b_0^1}{%
\sqrt{s}^3}\int d\sigma\, \sigma ^k- 
$$
$$
\int\!\!\int d\sigma d\sigma ^{\prime } \,
\left( \frac{\sigma +\sigma ^{\prime }}%
2\right) ^k2G^{\mu \nu }_s(\sigma ,\sigma ^{\prime })\left. (-g_{\gamma \mu }2%
{\cal D}z^{\prime \gamma })\right| _\sigma \frac{D\Psi }{Dz^\nu (\sigma
^{\prime })} \,\, ,
$$
while the third
$$
\left[ -\int d\sigma\, \bar \Delta {\cal \,,}\widetilde{{\cal B}}\right] \Psi
= 
$$
$$
-\int d\sigma\, \left[ (\bar \Delta {\cal \,}\widetilde{{\cal B}})\Psi
+2G^{\mu \nu }_s\frac{D\widetilde{{\cal B}}}{Dz^\mu }\frac{D\Psi }{Dz^\nu }%
\right] = 
$$
$$
0-\int\!\!\int d\sigma d\sigma ^{\prime }\,
2G^{\mu \nu }_s(\sigma ,\sigma ^{\prime
})\left( -2g_{\gamma \mu }k\sigma ^{k-1}z^{\prime \gamma }-2g_{\gamma \mu
}\sigma ^k{\cal D}z^{\prime \gamma }\right) \frac{D\Psi }{Dz^\nu (\sigma
^{\prime })} \,\, .
$$
The second and third terms together give 
$$
4\frac{k(k-1)}4\frac{b_0^0}{\sqrt{s}}N\int d\sigma\, \sigma
^{k-2}\Psi -4 \frac{2N}4\frac{b_0^1}{\sqrt{s}^3}\int d\sigma\, \sigma ^k\Psi
+4 \int d\sigma\, k\sigma ^{k-1}{\cal P}(\sigma )\Psi = 
$$
\be
\label{nana}
4\int d\sigma\, k\sigma ^{k-1}{\cal P}(\sigma )\Psi 
\ee
as the first and second terms vanish under the antisymmetry
of the commutator with respect to $(\frac{\sigma +\sigma ^{\prime }}2)^k$ 
and $1$.
The previous result is independent of the choice of the functional $\Psi $%
. However, the first term of (\ref{vise}) 
depends on the specific form of $\Psi $. We can
take $\Psi $, as in the previous section, to be the vacuum wave functional.
It is possible to construct the vacuum functional as a derivative 
expansion of its logarithm in local functionals (see \cite{Paul}, \cite{PJ}). 
Working in the lowest order of approximation for slowly varying fields
(small momentum) the
vacuum functional becomes $\Psi =e^{\int f_{\mu \nu }z^{\prime \mu
}z^{\prime \nu }}$, where $f_{\mu \nu }$ is an ultra-local function which
can be determined from the Schr\"odinger equation. The following
calculations will be performed for up to two derivatives 
with respect to $\sigma $.
For example, the action of the two functional derivatives on $\Psi $ gives
$$
\frac{\delta ^2}{\delta z(\sigma )\delta z(\sigma ^{\prime })}e^{\int f_{\mu
\nu }z^{\prime \mu }z^{\prime \nu }}=\frac{\delta ^2\int f_{\mu \nu
}z^{\prime \mu }z^{\prime \nu }}{\delta z(\sigma )\delta z(\sigma ^{\prime })%
}e^{\int f_{\mu \nu }z^{\prime \mu }z^{\prime \nu }}+\frac{\delta \int
f_{\mu \nu }z^{\prime \mu }z^{\prime \nu }}{\delta z(\sigma )}\frac{\delta
\int f_{\mu \nu }z^{\prime \mu }z^{\prime \nu }}{\delta z(\sigma ^{\prime })}%
e^{\int f_{\mu \nu }z^{\prime \mu }z^{\prime \nu }} 
$$
from which we only consider the first term, as the second involves four
derivatives with respect to $\sigma $. So we only need to consider the $\int
f_{\mu \nu }z^{\prime \mu }z^{\prime \nu }$ functional in the expansion of
the exponential $\Psi $. For this case we have 
$$
M^k\int f_{\mu \nu }\,z^{\prime \mu }z^{\prime \nu }= 
$$
$$
\int d\sigma\, \sigma ^k\left\{ (J_2f)_{\mu \nu }z^{\prime \mu }z^{\prime \nu
}+\frac 1sb_0^0f_{\,\,\nu }^\nu \right\} +\int d\sigma\, k\sigma
^{k-1}2b_0^0D^\nu f_{\nu \mu }z^{\prime \mu }+\int d\sigma\, \frac{k(k-1)}%
4\sigma ^{k-2}b_0^02f_{\,\,\nu }^\nu \,\, , 
$$
also
$$
\Delta M^k\int f_{\mu \nu }z^{\prime \mu }z^{\prime \nu }= 
$$
$$
..+\int d\sigma\, k\sigma ^{k-1}2b_0^0(J_1D^\nu f_\nu )_\mu z^{\prime \mu
}+\int d\sigma\, \frac{k(k-1)}4\sigma ^{k-2}b_0^02(J_0f_{\,\,\nu }^\nu ) 
$$
and
$$
M^k\Delta \int f_{\mu \nu }z^{\prime \mu }z^{\prime \nu }= 
$$
$$
..+\int d\sigma\, k\sigma ^{k-1}2b_0^0D^\nu (J_2f)_{\nu \mu }z^{\prime \mu
}+\int d\sigma\, k(k-1)\sigma ^{k-2}b2(J_2f)_{\,\,\nu }^\nu \,\, ,
$$
where 
\bq
\left( J_nf\right)_{\rho_1..\rho_n}
&=&
b^0_0 g^{\mu\nu}\left( D_\mu D_\nu\, f_{\rho_1..\rho_n}
+n\,f_{\lambda(\rho_2..\rho_n}R_{\rho_1)\mu\nu}^{\,\,\,\,\,\,\,
\lambda}\right)
\nonumber\\
&&
-n(n-1)\left(b_1^1g^{\mu\nu}\,f_{\mu\nu (\rho_3..\rho_n}
g_{\rho_1\rho_2)}
+b_2^1\,f_{\rho_1..\rho_n}\right),\label{eq:J}
\eq
and ``..'' represent homogeneous terms, which do not contribute to the 
commutator.
Using the identities
$$
(\text{tr }J_nf)_{\rho _3..\rho _n}-(J_{n-2}\text{tr }f)_{\rho _3..\rho n}=0 
$$
$$
(D\cdot J_nf)-J_{n-1}(D\cdot f))_{\rho _2..\rho _n}=0 
$$
from \cite{PJ}, when we asked for the Poincar\'e algebra to be satisfied 
($k = 1$), we get
$$
(\Delta M^k-M^k\Delta ) \int f_{\mu \nu }z^{\prime \mu
}z^{\prime \nu }=0 \,\, .
$$
So altogether for the vacuum functional $\Psi =e^{\int f_{\mu \nu }z^{\prime
\mu }z^{\prime \nu }}$ the commutator up to the first order becomes 
\begin{equation}
\label{comn}\frac 14\left[ \int d\sigma\, {\cal H,}\int d\sigma\, \sigma ^k{\cal %
H}\right] \Psi =\int d\sigma\, k\sigma ^{k-1}{\cal P}\Psi \,\, .
\end{equation}
No additional term results! This could be derived, more easily, by
substituting $f_{\mu \nu }=ag_{\mu \nu }$, which is the most general form the
first term could have, in the expansion of the logarithm of the vacuum
functional, for an appropriate constant $a$. From (\ref{comn}) we deduce
that in addition to the vacuum state, all the exited states will not
contribute to the commutator (\ref{vise}) at the first order as $f$ is general.
We can calculate the central charge term appearing 
when we use a second order test functional 
$F=\int f_{\mu\nu}{\cal D}z^{\prime \mu}{\cal D}z^{\prime \nu}$. 
The action of $M_k$ on $F$ is 
\bq
&&
M^k \int f_{\mu \nu}{\cal D}z^{\prime \mu}{\cal D}z^{\prime \nu}=
\nonumber\\
&&
\int d \sigma \, \sigma^k \Big\{ {{2b_0^2} \over \sqrt{s}^5} f_\nu ^\nu +
{{4b_0^1} \over \sqrt{s}^3} D^\nu f_{\nu \mu}{\cal D}z^{\prime \mu}+
{1 \over \sqrt{s}}(\bar J_2 f)_{\mu \nu}{\cal D}z^{\prime \mu}
{\cal D}z^{\prime \nu} \Big\}+
\nonumber\\
&&
\int d\sigma\, \sigma^{k-2} k(k-1)\Big\{
{b_0^0 \over \sqrt{s}} D^\nu f_{\nu \rho_2}{\cal D}z^{\prime \rho_2}+
{{2b_0^0} \over \sqrt{s}} f_{\kappa \nu}z^{\prime \rho}z^{\prime \gamma}
R^{\nu\,\,\,\,\,\,\kappa}_{\,\,\,\gamma\rho} -
{b_0^1 \over \sqrt{s}^3}f_\nu ^\nu \Big\}+
\nonumber\\
&&
\int d\sigma\, \sigma^{k-4} {k(k-1)(k-2)(k-3) \over 8} 
{b_0^0 \over \sqrt{s}} f_\nu ^\nu+ 
\int d\sigma\, k \sigma^{k-1} {{4b_0^0} \over \sqrt{s}}
R_\gamma ^{\,\,\,\rho} f_{\kappa \rho}
z^{\prime \gamma}{\cal D}z^{\prime \kappa} \,\, ,
\eq
where
\be
(\bar J_2 f)_{\mu\nu}=b_0^0 \bar \Delta f_{\mu\nu} +
2b_0^0 R_\mu ^\rho f_{\nu \rho} \,\, .
\ee
Using the relation
\be
M^k \int f_\rho {\cal D}z^{\prime \rho}=...+
\int d\sigma\, \Big\{ k \sigma ^{k-1} {{2b_0^0} \over \sqrt{s}}
R_\gamma ^\rho f_\rho z^{\prime \gamma}+ 
{k(k-1) \over 2}\sigma^{k-2} {b_0^0 \over \sqrt{s}} 
D^\nu f_\nu \Big\}
\ee
we can show that the commutator $(\Delta M^k -M^k \Delta)F$ is
\bq
&&
(\Delta M^k -M^k \Delta)F=
\nonumber\\
&&
\int d\sigma \, \sigma^{k-2} k(k-1) \Big\{
-{{2b_0^0 b_0^1} \over {s^2}} \Big[ D^\nu D^\mu f_{\mu\nu} +
R^{\nu \kappa} f_{\nu \kappa}\Big] +
\nonumber\\
&&
{{{b_0^0}^2 }\over s} 
\Big[ 2R_{\gamma \rho} D^\nu f_{\nu \rho} z^{\prime \gamma}-
{1 \over b_0^0} D^\nu (\bar J_2 f)_{\nu \rho} {\cal D}z^{\prime \rho} +
{1 \over {a^4}}(f_{\gamma \rho}+ (N-2)f_\nu ^\nu g_{\gamma \rho})
z^{\prime \gamma}z^{\prime \rho}\Big]
\Big\}-
\nonumber\\
&&
\int d\sigma\, \sigma^{k-4} {{k(k-1)(k-2)(k-3)} \over 4} {{b_0^0}^2 \over s}
R^{\nu \rho} f_{\nu \rho} \,\, .
\label{jojo}
\eq
This is non-zero even for $f_{\mu \nu}=g_{\mu\nu}$.

\section{Modification of the Central Charge Term}

Let us ignore regularization problems and define the 
generalized Virasoro operators as
$$
L[u]=\frac 14\int u(\sigma )(P^\mu P_\mu +z^{\prime \mu
}z_\mu ^{\prime }-z^{\prime \mu }P_\mu -P^\mu z_\mu ^{\prime })d\sigma  
$$
$$
\widetilde{L}[u]=\frac 14\int u(-\sigma )(P^\mu P_\mu
+z^{\prime \mu }z_\mu ^{\prime }+z^{\prime \mu }P_\mu +P^\mu z_\mu ^{\prime
})d\sigma \,\, ,  
$$
where the operator $\widetilde{L}$ is the same as $L$, but with $\sigma $
replaced by $-\sigma $. 
With these definitions we can express $H$, $L^k$ and $P$ in terms
of $L$ and $\widetilde{L}$ as%
$$
L[1]+\widetilde{L}[1]=\frac 12\int (P^\mu P_\mu +z^{\prime \mu }z_\mu
^{\prime })=H 
$$
$$
L[1]-\widetilde{L}[1]=-\frac 12\int (z^{\prime \mu }P_\mu -\frac 12z^{\prime
\mu }P_\mu )=-P 
$$
$$
L[\sigma ]+\widetilde{L}[\sigma ]=-\int \sigma z^{\prime \mu }P_\mu =-\int
\sigma {\cal P} 
$$
$$
L[\sigma ]-\widetilde{L}[\sigma ]=\frac 12\int \sigma (P^\mu P_\mu
+z^{\prime \mu }z_\mu ^{\prime })=L^1
$$
and so on.
Using these identifications we can see how an extension to the algebra (\ref
{comn}), will be placed in the modified Virasoro algebra. Let
us extend the algebra with the additional term $A$, as
$$
\left[ L[1],L[\sigma ^k]\right] =-L\left[ [1,\sigma ^k]\right] +A_k
$$
where
$$
[u,v]=uv^{\prime }-u^{\prime }v\Rightarrow [1,\sigma ^k]=k\sigma ^{k-1} 
$$
and
$$
\left[ \widetilde{L}[1],\widetilde{L}[\sigma ^k]\right] =-\widetilde{L}%
\left[ [1,\sigma ^k]\right] +\widetilde{A}_k 
$$
We separate two cases: $k=2l$ and $k=2l+1$. Then%
$$
L[\sigma ^{2l}]=\frac 14\int \sigma ^{2l}(P^2+z^{\prime 2}-z^{\prime
}P-Pz^{\prime }) 
$$
$$
\widetilde{L}[\sigma ^{2l}]=\frac 14\int \sigma ^{2l}(P^2+z^{\prime
2}+z^{\prime }P+Pz^{\prime }) 
$$
so that
$$
L[\sigma ^{2l}]+\widetilde{L}[\sigma ^{2l}]=\frac 12\int \sigma
^{2l}(P^2+z^{\prime 2})=\int \sigma ^{2l}{\cal H} 
$$
and similarly
$$
L[\sigma ^{2l+1}]=\frac 14\int \sigma ^{2l+1}(P^2+z^{\prime 2}-z^{\prime
}P-Pz^{\prime }) 
$$
$$
\widetilde{L}[\sigma ^{2l+1}]=-\frac 14\int \sigma ^{2l+1}(P^2+z^{\prime
2}+z^{\prime }P+Pz^{\prime }) 
$$
so that%
$$
L[\sigma ^{2l+1}]-\widetilde{L}[\sigma ^{2l+1}]=\frac 12\int \sigma
^{2l+1}(P^2+z^{\prime 2})=\int \sigma ^{2l+1}{\cal H} \,\, .
$$
For the first case we have%
$$
\left[ \int {\cal H,}\int \sigma ^{2l}{\cal H}\right] =\left[ L[1]+%
\widetilde{L}[1],L[\sigma ^{2l}]+\widetilde{L}[\sigma ^{2l}]\right] = 
$$
$$
-L\left[ [1,\sigma ^{2l}]\right] +A_{2l}-\widetilde{L}\left[ [1,\sigma
^{2l}]\right] +\widetilde{A}_{2l}= 
$$
$$
2l\int \sigma ^{2l-1}z^\mu P_\mu +A_{2l}+\widetilde{A}_{2l} \,\, ,
$$
while for the second%
$$
\left[ \int {\cal H,}\int \sigma ^{2l+1}{\cal H}\right] =\left[ L[1]+%
\widetilde{L}[1],L[\sigma ^{2l+1}]-\widetilde{L}[\sigma ^{2l+1}]\right] = 
$$
$$
-L\left[ [1,\sigma ^{2l+1}]\right] +A_{2l+1}+\widetilde{L}\left[ [1,\sigma
^{2l+1}]\right] -\widetilde{A}_{2l+1}= 
$$
$$
(2l+1)\int \sigma ^{2l}z^\mu P_\mu +A_{2l+1}-\widetilde{A}_{2l+1} \,\, .
$$
If we compare the outcome of the commutator $\left[ \int {\cal H,}\int
\sigma ^k{\cal H}\right] $ acting on a general functional, then we can
identify the quantities $A$ and $\widetilde{A}$. For the functional 
$\int f_{\mu \nu }z^{\prime \mu }z^{\prime \nu }$ it gives $A=\widetilde{%
A}=0$.
Also, if we set $\widetilde{A}_{2l}=A_{2l}$ and 
$\widetilde{A}_{2l+1}=-A_{2l+1}$ we have for the functional
$\int f_{\mu \nu}{\cal D}z^{\prime \mu }{\cal D} z^{\prime \nu }$ that
\bq
\widetilde{A}_{k}&=&A_{k}=
\nonumber\\
&&
{1 \over 2} 
\int d\sigma \, \sigma^{k-2} k(k-1) \Big\{
-{{2b_0^0 b_0^1} \over {s^2}} \Big[ D^\nu D^\mu f_{\mu\nu} +
R^{\nu \kappa} f_{\nu \kappa}\Big] +
\nonumber\\
&&
{{{b_0^0}^2 }\over s} 
\Big[ 2R_{\gamma \rho} D^\nu f_{\nu \rho} z^{\prime \gamma}-
{1 \over b_0^0} D^\nu (\bar J_2 f)_{\nu \rho} {\cal D}z^{\prime \rho} +
{1 \over {a^4}}(f_{\gamma \rho}+ (N-2)f_\nu ^\nu g_{\gamma \rho})
z^{\prime \gamma}z^{\prime \rho}\Big]
\Big\}-
\nonumber\\
&&
\frac 12
\int d\sigma\, \sigma^{k-4} {{k(k-1)(k-2)(k-3)} \over 4} {{b_0^0}^2 \over s}
R^{\nu \rho} f_{\nu \rho} \,\, .
\label{sisa}
\eq
Finally, we note that the central charge we arrived at in Section IV
was a result of the non locality of the vacuum state. This is described by
the function $H(\sigma ,\sigma ^{\prime })$, which in contrast to the usual
delta function (appearing when we use local functionals), brings in the
non-local character of the vacuum and gives the specific
form to the central charge term. 
However, as we have calculated, we get 
operator-like terms for the central charge of the algebra on $S^N$
which is a fingerprint of the generated mass. 

\section{A `Wrong' but Instructive Example}

Let us consider the algebra of the usual Virasoro operators, but
acting on massive states, rather than on massless states of a conformal 
theory. For this we will take the free, massive scalar field 
theory, while the Virasoro operators, $L[u]$, will be the ones defined 
for the string. We will choose $u(\sigma)=\sigma^k$ 
and for the state to be the vacuum wave functional
\be
\label{funa}
\Psi[z]=exp \, - \int d\sigma \, z \sqrt{-\partial^2+m^2} z\,\, .
\ee
Using the general result of equation (\ref{nana})
we are interested in the first term of 
(\ref{vise}) as the rest give the usual terms of the algebra apart from
the anomaly ones. For an expansion of the logarithm of (\ref{funa}) 
valid for slowly varying fields, z, of the form
\be
z\, \sqrt{-\partial^2+m^2}\, z=z\, m(1- {\frac {\partial^2}{2m^2}}-
{\frac {(\partial^2)^2}{8m^4}}+...)\,z
\ee
we have
\bq
&&
M^k\Psi[z]= \int\!\!\int d\sigma d\sigma ^{\prime }\, 
\left({{\sigma +\sigma ^\prime} \over 2}\right)^k
G_s(\sigma
,\sigma ^{\prime })\frac \delta{\delta z (\sigma )}
\frac \delta{\delta z (\sigma ^{\prime})} \Psi[z] =
\nonumber\\
&&
\int d\sigma \,
m^2 \left[\left( b_0^0 \sigma^k -b_0^0 \frac {k(k-1)}{4m^2}\sigma^{k-2}
-\frac {b_0^1}{s} \frac{\sigma^k}{m^2} 
-b_0^0 \frac {k(k-1)(k-2)(k-3)}{64m^4} \sigma^{k-4}\right.\right.
\nonumber\\
&&
\left.\left. -\frac 32 \frac {b_0^1}s \frac {k(k-1)}{4m^4} \sigma^k 
-\frac {b_0^2}{s^2} \frac {\sigma^k}{4m^4} \right) \, z^2
+b_0^0 \frac {\sigma^k}{4m^4} z ^{\prime \prime 2} +...\right] 
\Psi[z] \,+\, (\text{const.}) \times \Psi[z]
\eq
so that
\bq
&&
\Delta M^k\Psi[z]=
\int\!\!\int d\sigma d\sigma ^{\prime }\, 
G_s(\sigma
,\sigma ^{\prime })\frac \delta{\delta z (\sigma )}
\frac \delta{\delta z (\sigma ^{\prime})} \times
\nonumber\\
&&
\int\!\!\int d\sigma d\sigma ^{\prime }\, 
\left({{\sigma +\sigma ^\prime} \over 2}\right)^k
G_s(\sigma
,\sigma ^{\prime })\frac \delta{\delta z (\sigma )}
\frac \delta{\delta z (\sigma ^{\prime})}
\Psi[z]=
\nonumber\\
&&
\int d\sigma\,m^2\, \left[ 
2 \frac {b_0^0}{\sqrt s} \left( 
b_0^0 \sigma ^k -b_0^0 \frac {k(k-1)}{4m^2} \sigma^{k-2}
-\frac {b_0^1}{s} \frac {\sigma^k}{m^2}
-b_0^0 \frac {k(k-1)(k-2)(k-3)}{64m^4} \sigma^{k-4} \right. \right.
\nonumber\\
&&
\left.\left.
-\frac32 \frac {b_0^1}{s} \frac {k(k-1)}{4m^4} \sigma^{k-2}
-\frac {b_0^2}{s^2} \frac {\sigma^k}{4m^4} \right)+
+2 \frac {b_0^2 b_0^0}{\sqrt{s}^5} \frac {\sigma^k}{4m^4}+...\right]
\Psi[z]
\nonumber\\ \nonumber\\
&&
\,\,\,\,\,\,\,\,\,\,\,\,\,\,\,\,\,\,\,\,\,\,\,\,\,\,\,+\,
(\text{terms}\,\,\text{with}\,\,z) \times \Psi[z]
\eq
as well as
\bq
&&
M^k \Delta \Psi[z]=
\int\!\!\int d\sigma d\sigma ^{\prime } \,
\left({{\sigma +\sigma ^\prime} \over 2}\right)^k
G_s(\sigma
,\sigma ^{\prime })\frac \delta{\delta z (\sigma )}
\frac \delta{\delta z (\sigma ^{\prime})} \times
\nonumber\\
&&
\int\!\!\int d\sigma d\sigma ^{\prime }\, 
G_s(\sigma
,\sigma ^{\prime })\frac \delta{\delta z (\sigma )}
\frac \delta{\delta z (\sigma ^{\prime
})}
\Psi[z]=
\nonumber\\
&&
\int d\sigma\, m^2\, \left[ 
2 \frac {b_0^0}{\sqrt s}\left( b_0^0 -\frac {b_0^1}{m^2s} 
-\frac {b_0^2}{4m^4s^2} \right) \right.
\nonumber\\
&&
\left.
+ \frac {b_0^0}{2m^4} \left( \frac {b_0^0}{\sqrt{s}}
\frac {k(k-1)(k-2)(k-3)}{16} \sigma^{k-4}+
\frac32 \frac {b_0^1}{\sqrt {s}^3} k(k-1) \sigma^{k-2}
+\frac {b_0^2}{\sqrt {s}^5} \sigma^k \right) +... \right]
\Psi[z]
\nonumber\\ \nonumber\\
&&
\,\,\,\,\,\,\,\,\,\,\,\,\,\,\,\,\,\,\,\,\,\,\,\,\,\,\,
+\,(\text{terms}\,\,\text{with}\,\,z) \,\, \times \Psi[z].
\eq
Finally, their difference is
\bq
\nonumber\\
(\Delta M^k-M^k \Delta)\Psi[z]=&&
\nonumber\\ \nonumber\\
\int d\sigma\, m^2 && \left(
-{{{b_0^0}^2} \over {m^4 \sqrt {s}}} 
{ {k(k-1)(k-2)(k-3)} \over {16}} \sigma^{k-4}
-{3 \over 2}{{b_0^0 b_0^1} \over {m^4 \sqrt{s}^3}} k(k-1) \sigma ^{k-2}
\right.
\nonumber\\
&&
\left.
-{{{b_0^0}^2} \over {2m^2 \sqrt{s}}} k(k-1) \sigma^{k-2}
+...\right) \Psi[z]
\nonumber\\ \nonumber\\
&&
\!\!\!\!\!\!\!\!\!\!\!\!\!\!\!\!\!\!\!\!\!\!\!+\,
(\text{terms}\,\, \text{with} \, z) \,\, \times \Psi[z].
\label{kaka}
\eq
The integral $\int d\sigma \, \sigma^r$ diverges as $1/\sqrt {s}^{r+1}$,
for $r$ a positive integer, with a certain regularization.
Now we see how the 
mass appears explicitly. This, compared with relation (\ref{jojo}),
gives us a taste of the way we expect 
the mass to appear in the central charge term of the $O(N)$ $\sigma$-model.

The power series in $1/\sqrt{s}$ in (\ref{sisa}) or (\ref{kaka})
seems to suggest divergences of increasing order as $s \rightarrow 0$.
But the expansion of the vacuum functional and of the kernel, 
$G^{\mu \nu}_s$, is consistent for large $s$, which is equivalent 
to small momentum.
By analyticity arguments the series can be related to small $s$ with 
resummation procedures (see \cite{PMJ}) from where we can extract the 
correct divergence behavior of the central charge term.
The resumed series should give a 
result which is independent of the regularization procedure if we 
include enough terms in the power series. In other words, a different 
point splitting kernel satisfying the proper transformation properties 
and initial conditions, should give the same answer after resummation 
is used. Thus, in order to get an improved approximation we need more 
terms in the expansion of the kernel beyond the second order.
This can be achieved, for example, by the use of computer programs.

Finally, we saw above that in the lowest order in our expansion scheme 
the central charge, $A_k$, is zero. To find its exact form in the next 
order we need to know what $f_{\mu \nu}$ is. Apart from deriving 
$A_k$ by solving the Schr\"odinger equation for $f_{\mu \nu}$ 
(see \cite{PJ}), it can also be calculated from a renormalization 
group study of correlators of stress-energy 
tensor products \cite{anselmi}. Thus, knowing the particular form of 
the central charge term as given in (\ref{sisa}) we can extract the form of 
$f_{\mu \nu}$ without having to solve the Schr\"odinger equation,
resulting to the construction of the vacuum wave functional. In
addition we believe that having this form of the central charge in hand 
would make easier the exact evaluation of it with another method.


\begin{references}

\bibitem{Green} M. B. Green, J. H. Schwarz, E. Witten, ``Superstring Theory'',
Cambridge University Press 1987.

\bibitem{jackiw2} E. Benedict, R. Jackiw, H.-J. Lee, ``Functional
Schr\"odinger and BRST Quantization of $(1+1)$-Dimensional Gravity''
hep-th/9607062.

\bibitem{anselmi} D. Anselmi, ``Central Functions and Their Physical
Implications'', hep-th/9702056.

\bibitem{bollini} C. G. Bollini, M. C. Rocca, 
``Vacuum state of the quantum string without anomalies in any number 
of dimensions'', hep-th/9607023.

\bibitem{mana1} P. Mansfield, Annals of Physics, Vol. 180, No. 2, (1987)
330.

\bibitem{Paul} P. Mansfield, ``The Vacuum Functional at Large Distances'',
Phys. Lett. B 358 (1995) 287, hep-th/9508148.

\bibitem{PJ} P. Mansfield, J.Pachos,
``The $O(N)$ $\sigma$-model Laplacian'', 
Phys. Lett. B 365 (1996) 169, hep-th/9511041.

\bibitem{Jian} J. Pachos, ``Second Order Calculations of the 
$O(N)$ $\sigma$-model Laplacian'', 
Phys. Lett. B 394 (1997) 365, hep-th/9701074.

\bibitem{PMJ} P. Mansfield, J Pachos, M. Sampaio, ``Short Distance
Properties from Large Distance Behaviour'', hep-th/9702072.

\end{references}
\end{document}